\documentclass[twocolumn,prd,floatfix,preprintnumbers,nofootinbib,superscriptaddress]{revtex4-2}

\usepackage{bm}
\usepackage{times}
\usepackage{amssymb,amsbsy,amsmath,amsfonts}
\usepackage{graphicx}
\usepackage{float}
\usepackage{color}
\usepackage{morefloats}
\usepackage{rotating}
\usepackage{srcltx}
\usepackage{slashed}
\usepackage{subfigure}
\usepackage{multirow,diagbox}
\usepackage{verbatim}
\usepackage{tabularx}
\usepackage{tabularray}
\usepackage[outdir=./]{epstopdf}
\usepackage{epsfig}
\usepackage[normalem]{ulem}
\usepackage{hyperref}
\usepackage{tikz}
\usepackage{braket}
\hypersetup{colorlinks=true,linkcolor=blue,citecolor=blue,menucolor=black,urlcolor=black,filecolor=blue}
\usepackage{balance}
\usepackage{makecell}

\newcommand\bkthree[3]{\langle \vec{#1} \left| #2 \right| \vec{#3}\rangle}

\newcommand\cuti[2]{\theta(\Lambda_{#1}-#2)}

\newcommand\pli[2]{#2^{l_{#1}}}

\newcommand\rmd[0]{\mathrm{d}}
\newcommand\rmi[0]{\mathrm{i}}
\newcommand\rme[0]{\mathrm{e}}

\begin{document}

\title{Probing the hadronic molecular nature of the $\Omega(2012)$, $\Omega(2380)$, and $\Omega_c(3120)$ via femtoscopy correlation functions}

\author{Si-Wei Liu}~\email{liusiwei@impcas.ac.cn}
 \affiliation{State Key Laboratory of Heavy Ion Science and Technology, Institute of Modern Physics, Chinese Academy of Sciences, Lanzhou 730000, China} 
\affiliation{School of Nuclear Sciences and Technology, University of Chinese Academy of Sciences, Beijing 101408, China}
\author{Wen-Tao Lyu}
\affiliation{School of Physics, Zhengzhou University, Zhengzhou 450001, China}
\affiliation{Departamento de Física Teórica and IFIC, Centro Mixto Universidad de Valencia-CSIC Institutos de Investigación de Paterna, 46071 Valencia, Spain}
\author{Ju-Jun Xie}~\email{xiejujun@impcas.ac.cn}
\affiliation{State Key Laboratory of Heavy Ion Science and Technology, Institute of Modern Physics, Chinese Academy of Sciences, Lanzhou 730000, China} 
\affiliation{School of Nuclear Sciences and Technology, University of Chinese Academy of Sciences, Beijing 101408, China}
\affiliation{Southern Center for Nuclear-Science Theory (SCNT), Institute of Modern Physics, Chinese Academy of Sciences, Huizhou 516000, China}
\date{\today}

\begin{abstract}

We investigate the femtoscopic correlation functions of systems associated with the $\Omega(2012)$, $\Omega(2380)$, and $\Omega_c(3120)$ resonances, with the aim of elucidating their internal structures. By employing effective potential models that incorporate both $s$-wave and $d$-wave interactions, we calculate the correlation functions for the relevant coupled channels. Our numerical results reveal pronounced enhancement structures in the $\Xi^0K^-$ and $\Xi_c^+K^-$ correlation functions, which provide direct evidences for the dynamically generated $\Omega(2012)$ and $\Omega_c(3120)$ states. Furthermore, significant low-momentum enhancements are observed in the $\Xi^{*0}K^-$ and $\Xi^{*0}K^{*-}$ channel, which are attributed to the $\Omega(2012)$ and $\Omega(2380)$ resonances. These theoretical calculations provide crucial insights for future high-precision measurements at the LHC and RHIC, offering a novel and independent approach to determine the dynamically generated hadronic molecular nature of these $\Omega$ excited states.

\end{abstract}

\maketitle

\section{Introduction} \label{sec:Introduction}

Understanding the non-perturbative nature of Quantum Chromodynamics (QCD) in the low-energy regime remains a central challenge in hadron physics. A key manifestation of this challenge is the ``missing baryon'' problem, where numerous excited states predicted by the constituent quark model have not been experimentally observed~\cite{Capstick:1986ter,Loring:2001ky,Klempt:2009pi,Crede:2013kia,Pervin:2007wa,Faustov:2015eba,ParticleDataGroup:2024cfk}. This problem is particularly pronounced in baryon systems containing strange ($s$) and charm ($c$) quarks. For the $\Omega$ hyperon, composed of three strange quarks, the Particle Data Group (PDG) has historically listed only the ground state $\Omega(1672)$ and a few excited states, such as the $\Omega(2380)$ resonance~\cite{ParticleDataGroup:2024cfk}. 

Recently, significant experimental progress has been made in the spectroscopy of excited $\Omega$ and $\Omega_c$ baryons. In 2018, the Belle collaboration observed a new narrow $\Omega(2012)$ state in the $\Xi^0 K^-$ and $\Xi^- K_S^0$ invariant mass spectra~\cite{Belle:2018mqs}, which was subsequently confirmed by other experiments~\cite{Belle:2021gtf,Belle:2022mrg,BESIII:2024eqk,ALICE:2025atb}. Furthermore, in the heavy-flavor sector, a series of excited $\Omega_c$ states reported by the LHCb and Belle collaborations~\cite{LHCb:2017uwr,LHCb:2021ptx,LHCb:2023sxp,Belle:2017ext}, including the $\Omega_c(3120)$, has further expanded the frontier of baryon spectroscopy.

The internal structures of these newly discovered $\Omega(2012)$, $\Omega(2380)$ and $\Omega_c(3120)$ exotic states remain highly controversial and under extensive discussion. On the one hand, conventional quark models interpret these states as orbital excitations of the ground-state baryons~\cite{Xiao:2018pwe,Wang:2018hmi,Liu:2019wdr,Polyakov:2018mow,Arifi:2022ntc,Wang:2022zja,Zhong:2022cjx}. Calculations based on QCD sum rules also partially support this compact three-quark interpretation~\cite{Aliev:2016jnp,Aliev:2018syi,Aliev:2018yjo,Su:2024lzy}. On the other hand, given their proximity to the thresholds of specific hadron pairs, within chiral unitary approaches and meson-exchange models, these states can be described as hadronic molecules dynamically generated from meson-baryon interactions~\cite{Guo:2017jvc,Valderrama:2018bmv,Lin:2018nqd,Huang:2018wth,Pavao:2018xub,Lu:2020ste,Shen:2025xcq,Ikeno:2020vqv,Liu:2020yen,Song:2024ejc,Hofmann:2006qx,Sarkar:2004jh,Xie:2024wbd}. To distinguish between these two scenarios, experimental efforts have primarily relied on measurements of its decay branching fractions, such as the ratio of three- and two-body decays $\Omega(2012) \to \bar{K}\pi\Xi$ to $\bar{K}\Xi$~\cite{Belle:2022mrg,ParticleDataGroup:2024cfk}. On the theoretical side, within the molecular picture, the production of $\Omega(2012)$ resonance in the nonleptonic weak decays of $\Omega_c^0 \to \pi^+ \bar{K}\Xi(1530)$ $(\eta \Omega) \to \pi^+ (\bar{K}\Xi)^-$ and $\pi^+ (\bar{K}\Xi\pi)^-$ via final-state interactions of the $\bar{K}\Xi(1530)$ and $\eta \Omega$ pairs was investigated~\cite{Zeng:2020och,Ikeno:2022jpe}. These above theoretical results of Ref.~\cite{Zeng:2020och} were subsequently confirmed by the Belle Collaboration~\cite{Belle:2021gtf}. However, existing decay data are insufficient to reach a definitive conclusion due to the large uncertainties in the decay widths and branching fractions. Therefore, more experiments are urgently needed to accurately determine the nature of these $\Omega$ resonances. Crucially, the dynamical generations of states like the $\Omega(2012)$ and $\Omega_c(3120)$ involve non-trivial coupled-channel dynamics where $d$-wave interactions play an indispensable role~\cite{Lu:2020ste,Shen:2025xcq,Ikeno:2020vqv,Pavao:2018xub}.

In fact, the femtoscopy technique has emerged as a precise tool for probing hadron-hadron interactions and the nature of exotic candidates~\cite{ALICE:2020ibs,ALICE:2022yyh,Isshiki:2021bqh,CMS:2023jjt,ALICE:2011kmy,ALICE:2012aai,ALICE:2015hvw,ALICE:2015hav,ALICE:2019eol,ALICE:2018ysd,ALICE:2019buq,ALICE:2019gcn,ALICE:2020mfd,STAR:2014dcy,STAR:2018uho}. For example, the ALICE and STAR collaborations have successfully employed this technique to precisely measure the interactions of various systems, such as $pp$, $p\Lambda$, $\Lambda\Lambda$ and $p \Omega$~\cite{ALICE:2018ysd,ALICE:2019gcn,ALICE:2019eol,ALICE:2019hdt,ALICE:2019buq,STAR:2014dcy,Morita:2014kza,STAR:2018uho,ALICE:2020mfd}. Theoretically, numerous studies have also explored the interactions of $K^-p$, $K^+\bar K^0$, and $DK$, and other systems~\cite{Molina:2023jov,Liu:2023wfo,Liu:2023uly,Albaladejo:2023wmv,Liu:2024uxn,Liu:2025eqw,Xie:2026hpp}. These studies demonstrate the feasibility of investigating hadron-hadron interactions and the exotic states via correlation functions~\cite{Liu:2024uxn}. 

The femtoscopic correlation function for the $\Omega(2012)$ state was theoretically investigated in Ref.~\cite{Lin:2026ypf}, based on that the $\Omega(2012)$ is dynamically generated as a quasi-bound $\bar{K} \Xi(1530)$-$\eta \Omega$ molecular state, with its coupling to the $\bar{K} \Xi$ channel driven by $d$-wave transitions. Along this line, in the present work, we systematically investigate the femtoscopic correlation functions for the coupled-channel systems associated with the $\Omega(2012)$, $\Omega(2380)$ and $\Omega_c(3120)$. By employing the well-established effective potential models that incorporate both $s$-wave and $d$-wave interactions, we calculate the scattering amplitudes and the correlation functions for the relevant coupled channels.

The paper is organized as follows: Section~\ref{sec:interaction formalism} details the interaction potential and the parameters used to describe the $\Omega(2012)$, $\Omega(2380)$, and $\Omega_c(3120)$ resonances. Section~\ref{sec:correlation function} briefly reviews the theoretical formalism of correlation functions including higher partial wave contributions. Section~\ref{sec:Results} presents the numerical results and discussions. Finally, Section~\ref{sec:Summary and Conclusions} summarizes the paper.


\section{Formalism} \label{sec:Formalism}

\subsection{The dynamical properties in coupled channels} \label{sec:interaction formalism}

In this section, we outline the coupled-channel formalism for meson-baryon interactions, which gives rise to the dynamical generation of multiple $\Omega$ resonances. This formalism incorporates the $\Xi^{*0}K^-$, $\Xi^{*-}\bar{K}^0$, $\Omega\eta$, $\Xi^{0}K^-$ and $\Xi^{-}\bar{K}^0$ channels responsible for the dynamical generation of the $\Omega(2012)$ state~\cite{Lu:2020ste,Shen:2025xcq}; the $\Xi^{*0}K^{*-}$, $\Xi^{*-}\bar{K}^{*0}$, $\Omega\omega$ and $\Omega\phi$ channels for the $\Omega(2380)$ resonance~\cite{Li:2026wck}; and the $\Xi_c^{*+}K^-$, $\Xi_c^{*0}\bar{K}^0$, $\Omega_c^{*0}\eta$, $\Xi_c^{+}K^-$ and $\Xi_c^{0}\bar{K}^0$ channels for the $\Omega_c(3120)$ state~\cite{Ikeno:2023uzz}. Here, $\Xi^*$, $\Xi^*_c$, and $\Omega^*_c$ denote $\Xi(1530)$, $\Xi_c(2645)$, and $\Omega_c(2770)$, respectively, following the PDG~\cite{ParticleDataGroup:2024cfk}.

Following the formalism in Refs.~\cite{Liu:2026esv,Aceti:2012dd,Gamermann:2009uq,Yamagata-Sekihara:2010kpd}, the elements of the scattering amplitude matrix for these above coupled channels in momentum space can be calculated as follows,
\begin{eqnarray}
        \bkthree{q}{T_{ij}}{p} &=& (2l_{ij}+1){P_l}_{ij}(\cos\theta_{\hat{q}\hat{p}})T_{ij}(\vec{q},\vec p) \nonumber \\
        && \times \cuti{i}{q}\cuti{j}{p},         \label{equ:T_single}
\end{eqnarray}
with $T_{ij}(\vec{q},\vec p) = t_{ij}\pli{ii}{q}\pli{jj}{p}$ and $l_{ij}$ denotes the orbital angular momentum between $i$-th and $j$-th channels. $\vec{q}$ and $q$ represent the three-momentum and its magnitude, respectively. The Legendre polynomial of order $l_{ij}$ is denoted by $P_{l_{ij}}(\cos\theta_{\hat{q}\hat{p}})$. The step function $\cuti{i}{p}$ serves as a regulator with the cutoff momentum $\Lambda_i$. The two-body scattering amplitude $t_{ij}$ that is independent of the momentum, can be calculated algebraically via
\begin{equation}
        t=\frac{v}{1-v \tilde{G}},        \label{equ:T_algebraical}
\end{equation}
where the interaction potentials $v_{ij}$ for all charged channels are independent of the momentum, which are summarized in Table \ref{tab:Omega_potentials}. Here, the $F$ is given by
\begin{equation}    
        F=-\frac{1}{4f^2}\left(k_i^0+k_j^0\right),    \label{equ:F}
 \end{equation}
with $f= 93$ MeV the pion decay constant, and $k_i^0 = (s+m_i^2-M_i^2)/(2\sqrt{s})$ the meson energy of the $i$-th channel in the meson-baryon rest frame with invariant mass $\sqrt{s}$. $m_i$ and $M_i$ are the the meson mass and the baryon mass in the $i$-th channel.

\begin{table}[htbp]
	\begin{center}
    \caption{The interaction potential matrices for different coupled channels in the charged basis.}
    \label{tab:Omega_potentials}
    \vspace{0.5em}
    
    \text{(a) $\Xi^{*0}K^-$, $\Xi^{*-}\bar{K}^0$, $\Omega\eta$, $\Xi^{0}K^-$, and $\Xi^{-}\bar{K}^0$ channels.} \\
    \vspace{0.5em}
    \renewcommand\arraystretch{1.5}	
	\begin{tabular}{cccccc}
	\Xhline{1pt}
	& $\Xi^{*0}K^-$ & $\Xi^{*-}\bar{K}^0$ & $\Omega\eta$ & $\Xi^{0}K^-$ & $\Xi^{-}\bar{K}^0$ \\ 
    \hline
	$\Xi^{*0}K^-$  & $-F$ & $F$ & $-\frac{3}{\sqrt{2}}F$ & $\frac{1}{2}\alpha_1$ & $-\frac{1}{2}\alpha_1$ \\ 
    $\Xi^{*-}\bar{K}^0$ & & $-F$ & $-\frac{3}{\sqrt{2}}F$ & $\frac{1}{2}\alpha_1$ & $-\frac{1}{2}\alpha_1$ \\
    $\Omega\eta$ & & & $0$ & $-\frac{1}{\sqrt{2}}\beta_1$ & $\frac{1}{\sqrt{2}}\beta_1$ \\
	$\Xi^{0}K^-$ & & & & $0$ & $0$ \\   	
    $\Xi^{-}\bar{K}^0$ & & & & & $0$ \\ \Xhline{1pt}
	\end{tabular}
    
    \vspace{1em} 
    
    \text{(b) $\Xi^{*0}K^{*-}$, $\Xi^{*-}\bar{K}^{*0}$, $\Omega\omega$, and $\Omega\phi$ channels.} \\
    \vspace{0.5em}
	\begin{tabular}{ccccc}
	\Xhline{1pt}
	& $\Xi^{*0}K^{*-}$ & $\Xi^{*-}\bar{K}^{*0}$ & $\Omega\omega$ & $\Omega\phi$ \\ 
    \hline
	$\Xi^{*0}K^{*-}$ & $-F$ & $F$ & $-\sqrt{\frac{3}{2}}F$ & $\sqrt{3}F$ \\ 
    $\Xi^{*-}\bar{K}^{*0}$ & & $-F$ & $-\sqrt{\frac{3}{2}}F$ & $\sqrt{3}F$ \\
    $\Omega\omega$ & & & $0$ & $0$ \\
	$\Omega\phi$ & & & & $0$ \\ \Xhline{1pt}
	\end{tabular}

    \vspace{1em} 

    \text{(c) $\Xi_c^{*+}K^-$, $\Xi_c^{*0}\bar{K}^0$, $\Omega_c^{*0}\eta$, $\Xi_c^{+}K^-$, and $\Xi_c^{0}\bar{K}^0$ channels.} \\
    \vspace{0.5em}
    \begin{tabular}{cccccc}
		\Xhline{1pt}
		&         $\Xi_c^{*+}K^-$  & $\Xi_c^{*0}\bar{K}^0$  & $\Omega_c^{*0}\eta$  & $\Xi_c^{+}K^-$  & $\Xi_c^{0}\bar{K}^0$     \\ 
        \hline
	    $\Xi_c^{*+}K^-$  & $0$     & $F$  & $-\frac{2\sqrt{6}}{3}F$  & $\frac{1}{2}\alpha_2$ & $\frac{1}{2}\alpha_2$ \\ 
     $\Xi_c^{*0}\bar{K}^0$ &      & $0$  & $-\frac{2\sqrt{6}}{3}F$  & $\frac{1}{2}\alpha_2$ & $\frac{1}{2}\alpha_2$  \\
        $\Omega_c^{*0}\eta$    & & & $0$ & $-\frac{1}{\sqrt{2}}\beta_2$ & $-\frac{1}{\sqrt{2}}\beta_2$   \\
		$\Xi_c^{+}K^-$  &   &   &  & $0$ & $0$ \\   	
  $\Xi_c^{0}\bar{K}^0$ &   &   &  &  & $0$ \\ \Xhline{1pt}
		\end{tabular}
	\end{center}
\end{table}

In Tables~\ref{tab:Omega_potentials} (a) and \ref{tab:Omega_potentials} (c), $\alpha_1$ ($\alpha_2$) and $\beta_1$ ($\beta_2$) represent the free parameters associated with $d$-wave couplings, which are extracted by fitting to the experimental mass and total width of the $\Omega(2012)$ and $\Omega_c(3120)$ resonances.

In Eq.~\eqref{equ:T_algebraical}, the loop function $\tilde G_{ii}$ for two stable particles is
\begin{equation}
        \tilde{G}_{ii} (s) =\int^{\Lambda_i}_0 \rmd^3k~G_{ii}(k) k^{2l_{ii}},         \label{equ:dotG}
 \end{equation}
with
\begin{eqnarray}
        G_{ii}(k) &=& \frac{2 M_i}{(2\pi)^3}  \frac{\omega_{M_i}+\omega_{m_i}}{2\omega_{M_i} \omega_{m_i}\left[s-(\omega_{M_i}+\omega_{m_i})^2 + i\epsilon\right]}, 
  \end{eqnarray}
where $\omega_{M_i} = \sqrt{M^2_i + k^2}$ and $\omega_{m_i} = \sqrt{m^2_i + k^2}$ are the energies of baryon and meson in $i$-th channel, respectively. For these channels involved unstable particles, the loop function in Eq.~(\ref{equ:dotG}) should be modified to incorporate finite-width effects~\cite{Lu:2020ste,Xie:2007qt}. For this purpose, we rewrite the loop function as
\begin{equation}    
\hat G_{ii} (s) = \frac{\iint \rmd\tilde M^2~\rmd\tilde m^2~ \rho_{M,i}(\tilde M^2) \rho_{m,i}(\tilde m^2)  \tilde G_{ii}(s,\tilde M^2,\tilde m^2)}{\iint \rmd\tilde M^2~\rmd\tilde m^2 \rho_{M,i}(\tilde M^2) \rho_{m,i}(\tilde m^2)},   \label{equ:Ghat}
\end{equation}
where $\rho_{M,i}$ and $\rho_{m,i}$ are the spectral functions for the baryons and mesons in the coupled channels, respectively, which are typically parameterized by a normalized Breit-Wigner distribution,
\begin{equation}    
    \begin{aligned}
    \rho(\tilde M^2)=\frac{1}{\pi}\frac{M \tilde \Gamma}{(\tilde M^2-M^2)^2+(M \tilde \Gamma)^2},
\end{aligned}
\end{equation}
with $M$ and $\tilde M$ the nominal mass of the unstable resonance and the integration variable of the invariant mass, respectively. Here, $\tilde\Gamma$ is the energy-dependent width,
\begin{equation}    
    \begin{aligned}
    \tilde \Gamma(\tilde M^2)&=\Gamma\frac{M}{\tilde M}\sum_n R_{n}\left(\frac{p_n(\tilde M)}{p_n(M)}\right)^{2l_{n}+1},
\end{aligned}
\end{equation}
with $\Gamma$ the nominal total decay width, and the subscript $n$ runs over all possible decay channels. $R_n$ and $l_n$ represent the decay branching fraction and orbital angular momentum for the partial decay into the $n$-th channel, respectively. $p_n(\tilde M)$ represents the center-of-mass three-momentum, with $M_{n,1}$ and $M_{n,2}$ are the particle masses in the $n$-th decaying channel.

The integration ranges in Eq.~\eqref{equ:Ghat} are set to $[M-6\Gamma, M+6\Gamma]$~\cite{Lu:2020ste,Pavao:2018xub}, depending on the nominal mass and width of the resonance. It is worth noting that for a stable particle ($\Gamma \to 0$), the Breit-Wigner distribution becomes a Dirac delta function, $\lim_{\Gamma\to 0} \rho(\tilde M^2)=\delta(\tilde M^2-M^2)$.

Next, we investigate the properties of the resonances generated by the coupled-channel interactions. Since resonance poles lie on the complex $\sqrt{s}$ plane, the loop function $\tilde G$ for two stable particles must be analytically continued to the second Riemann sheet
\begin{equation}
    \begin{aligned}
        \tilde G^{II}_{ii}(s) & = \left\{\begin{array}{ll}
        \tilde  G_{ii}^{(I)}(s),  & \text{Re}[\sqrt{s}]<\left(m_{i}+M_{i}\right) \\
        \tilde  G_{ii}^{(II)}(s), & \text{Re}[\sqrt{s}] \geq\left(m_{i}+M_{i}\right)
    \end{array},\right.
    \label{equ:G-complex-plane}
    \end{aligned}
\end{equation}
where $\tilde G_{ii}^{(I)}(s)$ is the loop function as in Eq.~\eqref{equ:dotG} and 
\begin{equation}
    \begin{aligned}
    \tilde G_{ii}^{(II)}(s) = \tilde G_{ii}^{(I)}(s) + i\frac{2M_i}{4\pi\sqrt{s}}p_i^{2l_{ii}+1}.
    \label{equ:G-second-riemann}
    \end{aligned}
\end{equation}
Here $p_i$ is the three momentum of the $i$-th channel. To ensure the correct analytic continuation, the imaginary part of the three-momentum $p_i$ must remain positive. By incorporating the finite width effects of the propagator on the second Riemann sheet, the pole position can be determined from
\begin{equation}
        \left.\text{det}\left(1-v \hat {G}^{II}\right)\right|_{s=s_{p}}=0,
        \label{equ:detT}
\end{equation}
where the pole corresponds to the mass and half-width of the resonance: $\sqrt{s_p}=M_R-i \Gamma_R/2$. And in the vicinity of the pole $s_{p}$, the coupling constants of the resonances to the various channels are extracted via the residue of the scattering amplitude,
\begin{equation}
    \begin{aligned}
        g_ig_j=\lim_{s\to s_{p}}(\sqrt{s}-\sqrt{s_{p}})T^{II}_{ij}, \label{eq:coupling}
\end{aligned}
\end{equation}
where $g_{i}$ denotes the coupling constant of the resonance to the $i$th channel. Once the couplings are established, the corresponding partial decay widths can be computed.

\subsection{Femtoscopic correlation functions and wave functions for general partial wave in coupled channels} \label{sec:correlation function}

\begin{table*}[htbp]
\renewcommand{\arraystretch}{1.3}
\centering
\caption{Dynamical free parameters and production weights for the $\Omega(2012)$, $\Omega(2380)$, and $\Omega_c(3120)$ states. The parameter $\omega_{i1}$ represents the production weight of channel $1$ in the calculation of the correlation function for channel 
$i$, and similarly for other indices. The numbers in parentheses denote the uncertainties in the last two digits.}
\label{tab:properties}
\begin{tabular}{c|ccc|cccccc} 
\Xhline{1pt}
\multirow{2}{*}{Resonances}     & \multicolumn{3}{c|}{Dynamical Free Parameters}                                             & \multicolumn{6}{c}{Production Weights}                              \\ 
                                & $q_\text{max}~(\text{MeV})$       & $\alpha~(10^{-8}~\text{MeV}^{-3})$                             & $\beta~(10^{-8}~\text{MeV}^{-3})$                  & $i$-th channel     & $\omega_{i1}$ & $\omega_{i2}$ & $\omega_{i3}$ & $\omega_{i4}$& $\omega_{i5}$ \\ 
\hline
\multirow{5}{*}{$\Omega(2012)$} & \multirow{5}{*}{1000} & \multirow{5}{*}{$-11.29(05)$} & \multirow{5}{*}{34.14(15)} & 1: $\Xi^{*0}K^-$   & 1             & 0.95(19)          & 0.09(02)      &1.21(24) &1.17(23)      \\
                                &                      &                                     &                          & 2: $\Xi^{*-}\bar{K}^0$    & 1.04(21)          & 1             & 0.10(02)    &1.23(24) &1.19(24)       \\
                                &                      &                                     &                          & 3: $\Omega\eta$    & 2.83(58)          & 2.76(57)             & 1      &2.93(60) &2.84(58)      \\
                                &                      &                                     &                          & 4: $\Xi^{0}K^-$     & 0.18(04)          & 0.16(03)          & 0       &1 &0.92(18)       \\
                                &                      &                                     &                          & 5: $\Xi^{-}\bar{K}^0$     & 0.21(04)          & 0.19(04)          & 0         &1.06(21) &1       \\
\hline
\multirow{4}{*}{$\Omega(2380)$} & \multirow{4}{*}{575} & \multirow{4}{*}{\textbackslash{}}      & \multirow{4}{*}{\textbackslash{}}       & 1: $\Xi^{*0}K^{*-}$   & 1             & 0.93(19)             & 0.64(13)      & 0 &       \\
                                &                      &                                     &                          & 2:~$\Xi^{*-}\bar{K}^{*0}$               & 1.05(21)             & 1         &0.69(14)  & 0  &  \\
                                &                      &                                     &                          & 3:~$\Omega\omega$    & 1.36(28)             & 1.31(27)             & 1        & 0  &     \\
                                &                      &                                     &                          & 4:~$\Omega\phi$    & 4.18(88)             & 4.07(86)             & 3.36(72)          & 1 &    \\
\hline
\multirow{5}{*}{$\Omega_c(3120)$} & \multirow{5}{*}{1000} & \multirow{5}{*}{$-4.12(03)$} & \multirow{5}{*}{47.10(37)} & 1: $\Xi_c^{*+}K^-$   & 1             & 0.95(19)          & 0.09(02)        &1.34(26) & 1.29(25)   \\
                                &                      &                                     &                          & 2: $\Xi_c^{*0}\bar{K}^0$    & 1.04(21)          & 1             & 0.09(02)     &1.38(27) &1.32(26)       \\
                                &                      &                                     &                          & 3: $\Omega_c^{*0}\eta$    & 3.04(62)          & 2.97(61)             & 1       &3.38(69) &3.26(67)     \\
                                &                      &                                     &                          & 4: $\Xi_c^{+}K^-$     & 0.19(04)          & 0.17(03)          & 0       &1 &0.93(18)       \\
                                &                      &                                     &                          & 5: $\Xi_c^{0}\bar{K}^0$     & 0.22(04)          & 0.20(04)          & 0         &1.06(21) &1       \\
\Xhline{1pt}
\end{tabular}
\end{table*}

The theoretical femtoscopic correlation function for a hadron-hadron pair in the coupled channels is given by
\begin{equation}
    C_i(\vec{p}_i) =\sum_{S,L,J}\tilde{\omega}_{(S,L,J)}  \sum_j  \int \rmd^3r~ \omega_{ij} ~S_j(\vec r)~|\Psi_j(\vec r)|^2,
    \label{equ:cf_total}
\end{equation}
where $\tilde{\omega}_{(S,L,J)}$ represents the statistical weight factor, which is determined by the individual spins $S_1$ and $S_2$, the total spin $S$, the orbital angular momentum $L$, and the total angular momentum $J$ of the system. The explicit formula for $\tilde{\omega}_{(S,L,J)}$ is~\cite{Mihaylov:2018rva,Ge:2025put},
\begin{equation}
     \tilde{\omega}_{(S,L,J)}=\frac{2S+1}{(2S_1+1)(2S_2+1)}\frac{2J+1}{(2L+1)(2S+1)}.
\end{equation}
The parameter $\omega_{ij}$ denotes the production weight of the hadron pair in the $j$-th channel relative to that in the $i$-th channel during the collision. The specific values of these weights are evaluated using the VLC method detailed in the appendix of Ref.~\cite{Encarnacion:2024jge}, where the momentum limit is relaxed from $300$ MeV to $500$ MeV. The source function $S_j\left(\vec{r}\right)$ is modeled as a static spherical Gaussian distribution,
\begin{equation}
        S_j\left(\vec{r}\right)=\frac{1}{\left(4\pi R_j^2\right)^{3/2}}~\rme^{-\frac{r^2}{4R_j^2}},
\end{equation}
where $R_j$ is the source size for the $j$-th channel. In the absence of experimental constraints on these channels, the source sizes $R_j$ are set to $1$ fm.

Furthermore, the scattering wave function in Eq.~(\ref{equ:cf_total}) can be separated into its free and interacting components,
\begin{equation}
    \begin{aligned}
        \Psi_j(\vec r)=\delta_{ij}\left[\phi_j(\vec r)-\phi_{j,l_{ij}}(\vec r)\right]+\Psi_{j,l_{ij}}(\vec r),
        \label{equ:wf}
    \end{aligned}
\end{equation}
where $\delta_{ij}$ denotes the Kronecker delta function and $\phi_j(\vec r)=\rme^{\rmi\vec p_j \cdot\vec r}$ is the free wave function. $\phi_{j,l_{ij}}(\vec r)$ represents the free wave function projected onto the partial wave with angular momentum $l_{ij}$, and $\Psi_{j,l_{ij}}(\vec r)$ is the corresponding full scattering wave function for that partial wave. These wave functions are given by
\begin{equation}
    \begin{aligned}
        \phi_{j,l_{ij}}(\vec r)&=\rmi^{l_{ij}}(2l_{ij}+1){P_{l_{ij}}}(\cos\theta_{\hat{p_i}\hat{r}})j_{l_{ij}}(p_j r)\\
        \Psi_{j,l_{ij}}(\vec r)&=\rmi^{l_{ij}}(2l_{ij}+1){P_{l_{ij}}}(\cos\theta_{\hat{p_i}\hat{r}})\\
        &\quad \left[\delta_{ij}j_{l_{ij}}(p_j r)+R_{ij}(\vec p_i)\right],
        \label{equ:wf-partial}
    \end{aligned}
\end{equation}
with the radial function
\begin{equation}
        R_{ij}(\vec p_i)=\int^{\Lambda_j}_0 \rmd^3q~j_{l_{ij}}(qr)G_{jj}(q)T_{ij}(\vec q,\vec{p}_i),
        \label{equ:radial-function}
\end{equation}
where $j_{l_{ij}}(qr)$ is the $l_{ij}$-th order spherical Bessel function and the partial wave matrix $l_{ij}$ for these channels are defined as 
\begin{equation}
    \begin{aligned}
        l_{ij,1} &= l_{ij,3}  = \left\{\begin{array}{ll}
        0, & i\le 2~\text{and}~j\le 2 \\
        2, & \text { else }
    \end{array},\right.\\
    l_{ij,2} &= 0,
    \label{equ:lij}
    \end{aligned}
\end{equation}
where $l_{ij,1}$, $l_{ij,2}$, and $l_{ij,3}$ represent the partial wave matrices of the channels in Tables~\ref{tab:Omega_potentials} (a), (b), and (c), respectively.

Furthermore, it is important to note that for coupled channels involving a pair of charged hadrons, such as the $\Xi_c^{*+} K^-$ and $\Xi_c^+ K^-$ channels in the $\Omega_c(3120)$ system, the Coulomb interaction plays a crucial role at low relative momenta and should be taken into account. Hence, the total amplitude including both strong and Coulomb interactions ($T^c_{l_{ij}}$) can be expressed as (see more details in Refs.~\cite{Liu:2026esv,Encarnacion:2024jge,Torres-Rincon:2023qll,Kamiya:2019uiw}),
\begin{equation}
    \begin{aligned}
        T_{ij}(\vec{q}, \vec{p}_i) = t_{ij} q^{l_{jj}} p_i^{l_{ii}}  + \delta_{ij} T^c_{l_{ij}}(q, p_i).
    \label{equ:Tij-coulomb}
    \end{aligned}
\end{equation}
Correspondingly, the free wave function $\phi_j(\vec r)$ and its partial-wave projection $\phi_{j,l_{ij}}(\vec r)$ in Eqs.~(\ref{equ:wf}) and (\ref{equ:wf-partial}) are replaced by the full Coulomb wave function and its partial-wave counterpart, respectively.

Finally, the generalized femtoscopic correlation function for arbitrary partial waves can be formulated as
\begin{equation}
    \begin{aligned}
    C_i(\vec{p}_i)&=\int \rmd^3r~ S_i(\vec r)\left|\tilde\phi_i(\vec r)-\tilde\phi_{i,l_{ii}}(\vec r)+\tilde j_{l_{ii}}(p_i r)\right|^2   \\
    &+\sum_{S,L,J} \sum_j (2l_{ij}+1)\tilde\omega_{(S,L,J)}\omega_{ij} \int \rmd^3r~  S_j(\vec r)  \\
    &~~~~\left[\left|R_{ij}(\vec p_i)\right|^2 +\delta_{ij}2\text{Re}\left[j_{l_{ij}}(p_j r) R_{ij}(\vec p_i)\right]\right] ,
    \end{aligned}
\end{equation}
with $\tilde j_{l_{ii}}(p_i r)=\rmi^{l_{ii}}(2l_{ii}+1){P_{l_{ii}}}(\cos\theta_{\hat{p_i}\hat{r}})j_{l_{ii}}(p_i r)$.

\section{Numerical results and discussions} \label{sec:Results}

In this section, we present the numerical results of coupled-channel interactions related to the $\Omega(2012)$, $\Omega(2380)$, and $\Omega_c(3120)$ states, based on the theoretical formalism constructed in the previous sections. Specifically, we identify the pole positions of these resonances by locating the singularities of the scattering amplitudes, and further compute the corresponding femtoscopic correlation functions to reveal their underlying dynamical characteristics.

\begin{table*}
\renewcommand{\arraystretch}{1.3}
\centering
\caption{The coupling constants ($g$) and decay branching fractions ($\mathcal{B}r$) of the $\Omega(2012)$ to the involving channels. Both isospin conserving (IC) and isospin breaking (IB) scenarios are considered. For the three-body decay channels, the couplings in the first two and the subsequent two columns correspond to the $\Omega(2012)\to \Xi^{*0}K^-$ and $\Omega(2012)\to\Xi^{*-}\bar K^0$ transitions, respectively.}
\label{tab:cp_bf}
\begin{tabular}{ccccccccc} 
\Xhline{1pt}
           \multicolumn{2}{c}{Channels}                                                   & $\Xi^- \pi^+ K^-$ & $\Xi^0 \pi^0 K^-$   & $\Xi^- \pi^0 \bar K^0$ & $\Xi^0 \pi^- \bar K^0$ & $\Xi^0 K^-$         & $\Xi^- \bar K^0$     & $\Omega\eta$  \\ 
\hline
\multirow{2}{*}{\begin{tabular}[c]{@{}c@{}}IB\end{tabular}}    & \begin{tabular}[c]{@{}c@{}}$\mathcal{B}r$\end{tabular} & 0.16              & 0.10                & 0.05                   & 0.10                   & 0.29                & 0.26                 & 0             \\
           & $g$                                                    & \multicolumn{2}{c}{$(1.49 + i0.02)$} & \multicolumn{2}{c}{$(1.53 + i0.02)$}         & $(0.15 + i0.02)$ & $(-0.14 - i0.02)$ & $-0.93$       \\
\multirow{2}{*}{\begin{tabular}[c]{@{}c@{}}IC\end{tabular}} & \begin{tabular}[c]{@{}c@{}}$\mathcal{B}r$\end{tabular} & 0.13              & 0.07                & 0.07                   & 0.13                   & 0.28                & 0.28                 & 0             \\
           & $g$                                                    & \multicolumn{2}{c}{$(1.51+0.02\text{i})$} & \multicolumn{2}{c}{$(1.51 +i 0.02)$}         & $(0.15 + i0.02)$ & $(-0.15 - i0.02)$ & $-0.93$       \\
\Xhline{1pt}
\end{tabular}
\end{table*}

In principle, the cutoff parameter $\Lambda_i$ in the loop function should take different values for different channels. Nevertheless, to reduce the number of free parameters, the cutoff momenta for all channels are taken as the same value as $\Lambda_i = q_{\rm max}$~\cite{Huang:2018wth,Pavao:2018xub,Lu:2020ste,Shen:2025xcq}. Specifically, they are fixed at $q_{\max}=1000~\mathrm{MeV}$ for the $\Omega(2012)$ and $\Omega_c(3120)$ systems, while for the $\Omega(2380)$ system we adopt $q_{\max}=575~\mathrm{MeV}$, which is consistent with Ref.~\cite{Li:2026wck}. With these cutoff parameters fixed, the remaining free parameters of the $\Omega(2012)$ system are constrained by the experimental mass, total decay width, branching fraction $\mathcal{B}r_{\Omega(2012)\to \Xi^0K^-}$~\cite{ALICE:2025atb}, and the branching-fraction ratio $\mathcal{R} = \mathcal{B}r_{\Omega(2012)\to \bar{K} \pi \Xi}/\mathcal{B}r_{\Omega(2012)\to \bar{K}\Xi}$~\cite{Belle:2022mrg}. For the $\Omega_c(3120)$ system, its dynamical parameters are determined exclusively from the measured mass and total decay width. The fitted parameters and production weights are summarized in Table~\ref{tab:properties}, and further details regarding these production weights can be found in Ref.~\cite{Encarnacion:2024jge}.

Based on these determined model parameters, the pole positions of the $\Omega(2012)$, $\Omega(2380)$, and $\Omega_c(3120)$ states are extracted in the complex energy plane and presented as follows
\begin{equation}
\begin{aligned}
\sqrt{s_{\Omega(2012)}} &= (2012.68 - 1.62i) ~{\rm MeV}, \\
\sqrt{s_{\Omega(2380)}} &= (2388.79 - 6.93i) ~{\rm MeV}, \\
\sqrt{s_{\Omega_c(3120)}} &= (3118.98 - 0.30i) ~{\rm MeV}.
\end{aligned}
\end{equation}

In addition, the couplings determined at the pole position with Eq.~\eqref{eq:coupling} and decay branching fractions of the $\Omega(2012)$ to individual channels can be evaluated (see more details in Ref.~\cite{Xie:2024wbd}), with the results summarized in Table~\ref{tab:cp_bf}. Consequently, the ratio of the three-body to two-body decay branching fractions $\mathcal{R}$ is evaluated to be $0.75$ in the isospin breaking (IB) scenario and $0.71$ in the isospin conserving (IC) scenario~\footnote{Isospin breaking refers to the case where the physical mass differences within the isospin multiplets in the final states are taken into account. By contrast, isospin conservation corresponds to calculations that neglect isospin breaking effects, adopting the average mass of the different charge states for the final-state isospin multiplets.}. Compared to the experimental values $\mathcal{B}r_{\Omega(2012)\to \Xi^0K^-}=0.34^{+0.16}_{-0.12}$, $\mathcal{B}r_{\Omega(2012)\to \Xi^- \bar{K}^0}=0.28^{+0.12}_{-0.07}$, and $\mathcal{R}=0.99\pm 0.26\pm 0.06$~\cite{ALICE:2025atb,Belle:2022mrg}, our theoretical calculations are consistent with the experimental measurements within the uncertainties. Moreover, the IB effects exhibit the same trend as those reported in Ref.~\cite{Zhong:2022cjx}, which were obtained within a chiral quark model. This is predominantly driven by the phase space because the $\Omega(2012)$ mass is close to the $\bar K \Xi(1530)$ mass threshold. For the channels involving $K^-$, its smaller mass enlarges the available phase space, yielding larger branching fractions in the IB case than in the IC one. Conversely, the opposite effect is observed for these channels involving $\bar K^0$.

\begin{figure*}[htbp]
\centering
\subfigure[The femtoscopic correlation functions for $\Xi^{*0}K^-$, $\Xi^{*-}\bar{K}^0$, $\Omega\eta$, $\Xi^{0}K^-$ and $\Xi^{-}\bar{K}^0$ channels]{\includegraphics[scale=0.35]{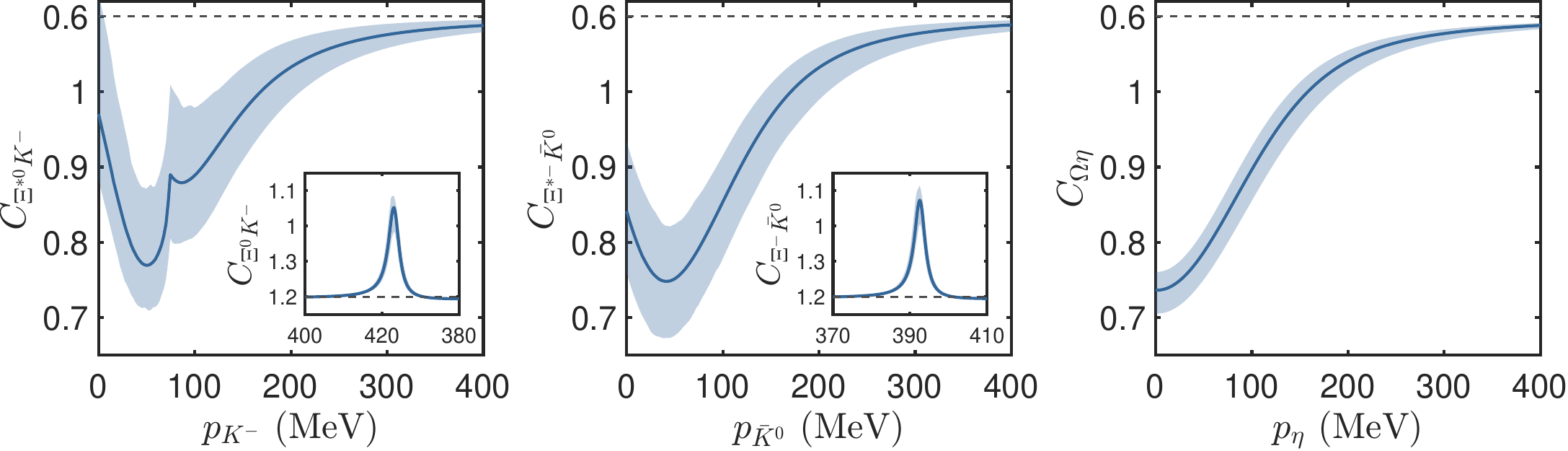} \label{fig:cf_KbarXi}}
\subfigure[The femtoscopic correlation functions for $\Xi^{*0}K^{*-}$, $\Xi^{*-}\bar{K}^{*0}$ and $\Omega\omega$ channels]{\includegraphics[scale=0.35]{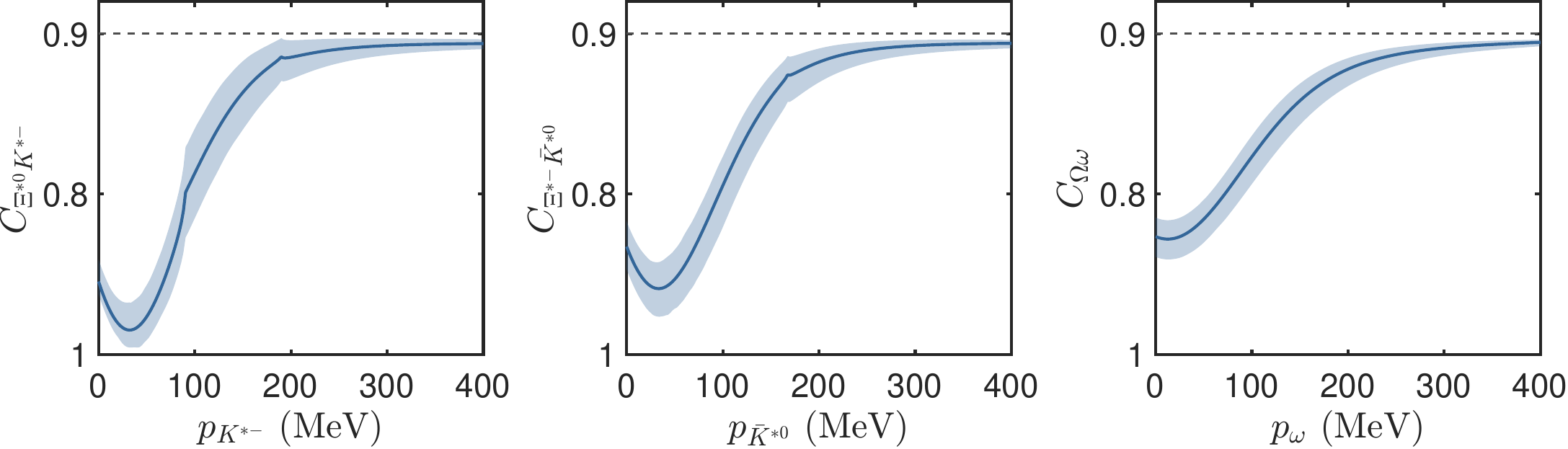} \label{fig:cf_KbarsXis}}
\subfigure[The femtoscopic correlation functions for $\Xi_c^{*+}K^-$, $\Xi_c^{*0}\bar{K}^0$, $\Omega_c^{*0}\eta$, $\Xi_c^{+}K^-$ and $\Xi_c^{0}\bar{K}^0$ channels]{\includegraphics[scale=0.35]{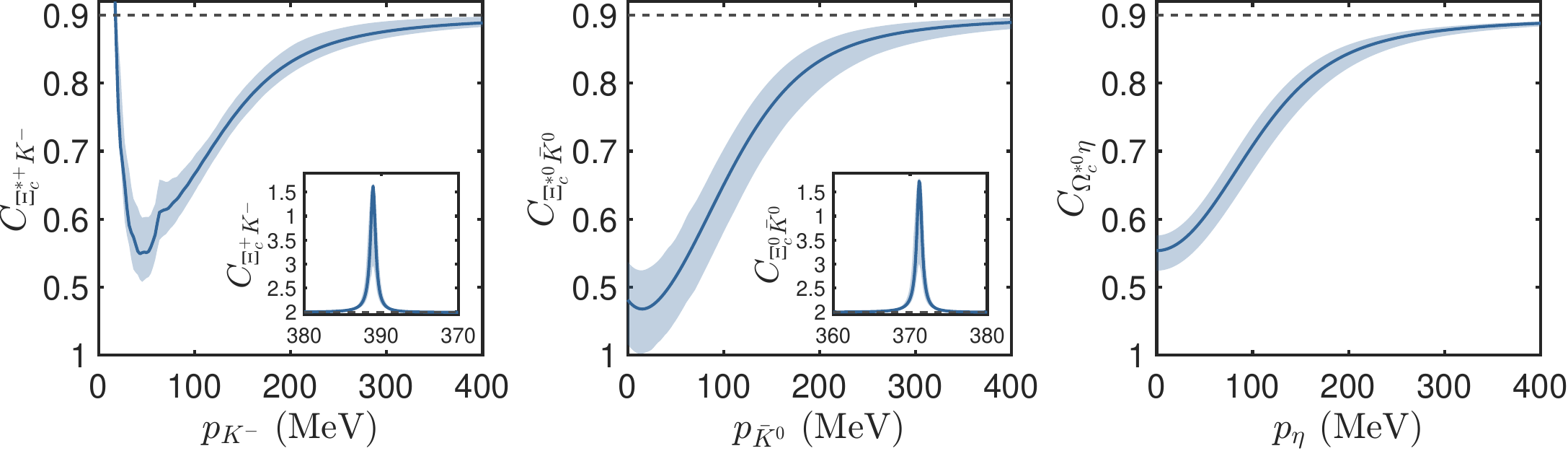} \label{fig:cf_KbarXic}}
\caption{The calculated correlation functions for the relevant channels. The grey bands represent the theoretical uncertainties, which incorporate the uncertainties of the dynamical free parameters and the production weights $\omega_{ij}$, and a $10\%$ variation in the source size $R_i$. The x-axis denotes the center-of-mass momentum within each specific system.}
\label{fig:cf}
\end{figure*}

With the scattering amplitudes and wave functions fully determined, the predicted femtoscopic correlation functions for all relevant channels are comprehensively presented in Fig.~\ref{fig:cf}. In these plots, the blue solid curves represent the central values of the correlation functions, while the shaded bands indicate the corresponding theoretical uncertainties. These bands include the uncertainties arising from the dynamical free parameters, the production weights $\omega_{ij}$, and a $10\%$ variation in the source size $R_i$.

As illustrated in Fig.~\ref{fig:cf}, pronounced enhancement structures are observed in the correlation functions of the $\Xi^0 K^-$, $\Xi^- \bar K^0$, $\Xi_c^+ K^-$, and $\Xi_c^0 \bar K^0$ channels. These structures can be attributed to the $\Omega(2012)$ and $\Omega_c(3120)$ states which are dynamically generated through coupled-channel interactions. These channels can therefore be regarded as the ``golden channels'' for experimentally probing the properties of these resonances. Furthermore, as pointed out in Ref.~\cite{Li:2026wck}, the $\Omega(2380)$ state can decay into the $\Xi^{*0}K^-$ and $\Xi^{*-}\bar K^0$ channels via box diagrams. Consequently, clear signatures of the $\Omega(2380)$ are expected to appear in the high-momentum regions of the corresponding correlation functions, providing a complementary means for its experimental identification.

Apart from the direct resonance peaks, the effects of these dynamically generated states can also manifest themselves as enhancements in the low-momentum region, since the pole masses of the $\Omega(2012)$ and $\Omega(2380)$ lie below the thresholds of the $\Xi^{*0}K^-$ and $\Xi^{*0}K^{*-}$ channels, respectively. Therefore, the threshold behavior is expected to be sensitive to their masses and decay widths, providing a useful indirect probe of the properties of these states. A similar threshold enhancement is theoretically expected in the $\Xi_c^{*+}K^-$ channel associated with the $\Omega_c(3120)$, but it is completely masked by the Coulomb attraction. This masking effect makes it practically impossible to extract the properties of the $\Omega_c(3120)$ resonance from the threshold behavior of this charged channel. In addition to the threshold enhancements, some cusp structures are observed in the correlation functions of the $\Xi^{*0} K^-$, $\Xi^{*0} K^{*-}$, $\Xi^{*-} \bar K^{*0}$, and $\Xi_c^+ K^-$. Such cusps serve as the signatures for the opening of higher-energy coupled channels, inherently reflecting the contributions of the coupled-channel interactions that can be measured in experiments.

\section{Summary and Conclusions} \label{sec:Summary and Conclusions}

We have systematically studied the femtoscopic correlation functions of the coupled-channel systems relevant to the $\Omega(2012)$, $\Omega(2380)$, and $\Omega_c(3120)$ resonances. By employing an effective potential model that incorporates higher partial-wave interactions, we extracted the corresponding pole positions and scattering amplitudes. Within this theoretical framework, we compute the femtoscopic correlation functions explicitly for the specific charge channels responsible for the dynamical generation of these states: the $\Xi^{*0}K^-$, $\Xi^{*-}\bar{K}^0$, $\Omega\eta$, $\Xi^{0}K^-$ and $\Xi^{-}\bar{K}^0$ channels for the $\Omega(2012)$ state; the $\Xi^{*0}K^{*-}$, $\Xi^{*-}\bar{K}^{*0}$ and $\Omega\omega$ channels for the $\Omega(2380)$ resonance; and the $\Xi_c^{*+}K^-$, $\Xi_c^{*0}\bar{K}^0$, $\Omega_c^{*0}\eta$, $\Xi_c^{+}K^-$ and $\Xi_c^{0}\bar{K}^0$ channels for the $\Omega_c(3120)$ state. The resulting correlation functions exhibit clear resonant structures in the $\Xi^0 K^-$ and $\Xi_c^+ K^-$ channels, which serve as characteristic signatures of the $\Omega(2012)$ and $\Omega_c(3120)$. Moreover, the low-momentum behavior in the $\Xi^{*0} K^-$ and $\Xi^{*0} K^{*-}$ channels is found to be sensitive to the masses and widths of these resonances.

These findings demonstrate that femtoscopic correlation measurements serve as powerful probes for extracting resonance properties, complementing conventional invariant mass spectrum analyses. High-precision femtoscopy measurements at the LHC and RHIC will be important for constraining the properties of these exotic states. In particular, high-precision measurements of their decay widths and branching fractions are indispensable for distinguishing between dynamically generated hadronic molecular configurations and conventional baryon structures.

\section*{Acknowledgments}

We acknowledge Pablo Encarnaci{\'o}n and Pan-Pan Shi for fruitful discussions. This work is partly supported by the National Key R\&D Program of China (Grant Nos. 2023YFA1606703 and 2024YFE0105200), the National Natural Science Foundation of China (Grant Nos. 12575094, 12435007, 12361141819, and 12475086), the Natural Science Foundation of Henan (Grant No. 252300423951), and the Zhengzhou University Young Student Basic Research Projects for PhD students (Grant No. ZDBJ202522). Wen-Tao Lyu acknowledges the support of the China Scholarship Council.

\bibliography{reference}

\end{document}